\documentclass{optica-article}

\journal{opticajournal} 

\articletype{Research Article}

\usepackage{lineno}

\usepackage{siunitx}
\usepackage[utf8]{inputenc} 
\usepackage[toc,page]{appendix}
\usepackage{amsmath} 
\usepackage{gensymb} 
\usepackage{threeparttable} 
\usepackage{booktabs} 
\usepackage{miller} 
\usepackage{xcolor}

\begin{document}

\title{A photonic integrated circuit for heterogeneous second harmonic generation}

\author{Theodore J. Morin\authormark{1,*}, Mingxiao Li\authormark{1}, Federico Camponeschi\authormark{1}, Hou Xiong\authormark{1}, Deven Tseng\authormark{1}, John E. Bowers\authormark{1}}

\address{\authormark{1}Department of Electrical and Computer Engineering, University of California, Santa Barbara, Santa Barbara, California 93106, USA}

\email{\authormark{*}tmorin@ucsb.edu} 

\begin{abstract*} 
Heterogeneous integration of GaAs-based lasers with frequency doubling waveguides presents a clear path to scalable coherent sources in the so-called green gap, yet frequency doubling systems have so far relied on separately manufactured lasers to deliver enough power for second harmonic generation. In this work, we propose a photonic integrated circuit (PIC) which alleviates the performance requirements for integrated frequency doublers. Two gain sections are connected by waveguides, with a frequency converter and a wavelength separator in between. The fundamental light circulates between the gain sections until it is converted and emitted through the wavelength separator. Variants of this separated gain PIC are discussed, and the PIC is implemented with thin film lithium niobate and directly bonded GaAs-based lasers, coupled by on-chip facets and adiabatic tapers, realizing visible light generation in the 515-595 nm range. 

\end{abstract*}

\section{Introduction}

It is notoriously difficult to produce light in the 520-600 nm range using semiconductor lasers \cite{Lu2024}. To support applications such as Na spectroscopy \cite{Huo2020}, light in this wavelength range is sometimes produced from longer wavelengths by using a nonlinear frequency doubler and mature GaAs-based lasers in the 1040-1200nm range \cite{Paschke:15}. In recent years, the proliferation of thin film lithium niobate (TFLN) \cite{Zhu2021, Luo2017, Luo2018} and recent results using the photogalvanic effect in Si$_3$N$_4$\cite{Lu2020, Li2023} have paved the way for increasingly compact and efficient systems for second harmonic generation (SHG). Further, coupling schemes based on on-chip facets plus adiabatic tapers (OFATs) offer a means to heterogeneously integrate lasers and waveguides despite large differences in material index \cite{Park2020, Morin2024rba}. Together, these advances suggest that heterogeneous integration can be used to streamline the production of green-yellow-orange coherent light sources.

An obstacle for such heterogeneous systems, however, is matching the performance of monolithic lasers. Although heterogeneous lasers have demonstrated excellent linewidth \cite{Xiang2023} and tunability \cite{tran2022extending}, power levels are typically lower than those afforded by the external light sources used in many frequency doubling demonstrations \cite{wang2024recent}. While this discrepancy is partly explained by the greater maturity of external laser fabrication processes, heterogeneous lasers must contend with inherent, nontrivial disadvantages. For instance, the passive layer of a typical heterogeneous platform places a thermal barrier between the laser and its heat sink \cite{Kaur2021}, which can limit output power as compared to cooler monolithic devices. For nonlinear applications in particular, overcoming this limitation is essential.

To address this challenge, in addition to improving the general performance of on-chip lasers, heterogeneous systems must make use of their inherent advantage: integration.

In this work, we propose a separated gain photonic integrated circuit (PIC) which can substantially increase the in-waveguide power offered by a given gain cross-section for the application of driving frequency up-conversion. The PIC is implemented with three different co-processed GaAs-based epitaxy designs, and measurable second harmonic generation is observed at green, orange and yellow wavelengths, despite unusually poor TFLN waveguide performance.

\section{Second harmonic generation PIC} \label{sec:SHGPIC}

\begin{figure}[htb]
	\centering
		\includegraphics[width=1.0\textwidth, draft=False]{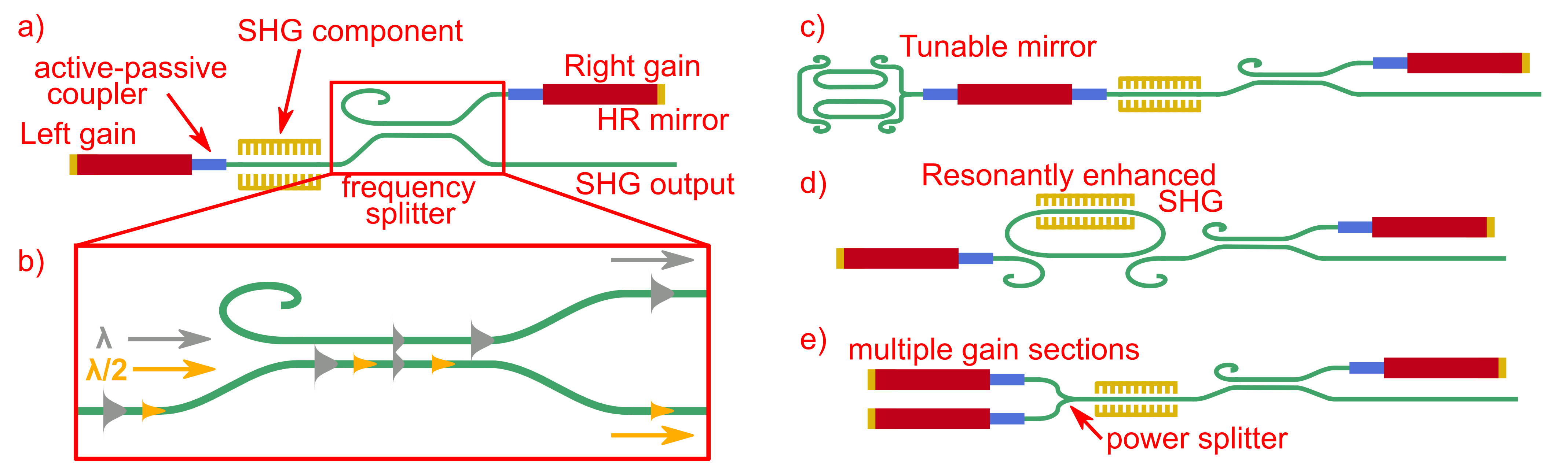}
	\caption[figshorttitle]{Photonic integrated circuits (PICs) for second harmonic generation (SHG). (a) Basic separated gain PIC. (b) Directional coupler for wavelength separation. (c) Multi-gain section PIC. (d) Resonant SHG PIC. (e) Vernier ring PIC.}
	\label{fig:CONCEPT}
\end{figure}

The proposed PIC is shown in Fig. \ref{fig:CONCEPT}a: Two gain sections point towards each other, with high reflectivity mirrors on their far ends. At the nearer ends, the two gain sections are coupled into two TFLN waveguides. These waveguides are coupled to each other with a wavelength separator which transfers light at the gain wavelength but not at the second harmonic wavelength; ideally all fundamental light is transferred between the two waveguides while no second harmonic light is transferred, as shown in Fig \ref{fig:CONCEPT}b. On one of the waveguides, a frequency doubling structure is included (in particular, a periodically poled lithium niobate waveguide). Because light converted in the frequency doubler is not transferred between the waveguides, it leaves the system as output. When the two gain sections are pumped sufficiently, the PIC lases at the gain wavelength, but its output is second harmonic light.

Several qualitative observations are worthwhile: 
\begin{itemize}
\item The wavelength dependent coupling between the two waveguides can be conveniently achieved with a directional coupler; 100\% directional couplers are robust against waveguide variations up to first order, and a suitable gap to transfer the gain wavelength will not transfer the more tightly confined second harmonic light. \item Since the two gain sections pump one another through the passive section, modes strongly coupled to the passive section will lase preferentially over undesirable modes with weaker coupling. At the same time, reflections between different parts of the cavity can offer longitudinal mode selection.
\item With sufficiently separation, each gain section can accept a large drive current without heating the other(s) and causing premature rollover. Thus, separating the gain sections can directly address the thermal limitations on output power from heterogeneous systems.
\item Because it requires minimal bending, the passive layout can easily avoid losses associated with tight bends, making it compatible with low-loss, weakly-guided waveguide platforms such as those described in \cite{Jin2021} and \cite{Churaev2023}.
    
\end{itemize}

Many variations are possible which may improve in-waveguide power, conversion efficiency, spectral performance, tuning or other properties. For instance, additional gain section(s) can be connected with integrated splitters (Fig. \ref{fig:CONCEPT}c); this configuration can increase the power in the united waveguide, while injection from the opposite gain section forces the two waveguides to lase in a common mode. A frequency doubling resonator could be included between the two gain sections to increase frequency conversion efficiency while also providing mode selectivity (Fig. \ref{fig:CONCEPT}d). To improve linewidth and frequency selection, one of the high reflectivity mirrors could be replaced with a Vernier ring mirror \cite{Tran2019} (Fig. \ref{fig:CONCEPT}e). In general, either or both of the gain sections may be replaced with any kind of laser, allowing a range of possible cavity designs.

\section{Fabrication} \label{sec:IMPL}

\cite{morin2024coprocessed} describes a platform for heterogeneous GaAs-based lasers on TFLN, and the same process was used to implement separated gain PICs, with only a few modifications: The waveguide layer was 200 nm of lithium niobate which was fully etched, rather than a partially etched 400-nm film. The bottom oxide was 2 $\mu$m rather than 4.7 $\mu$m. Before the waveguide etch, alignment marks were etched, and nickel electrodes were deposited and pulsed to form periodically poled lithium niobate (PPLN) for frequency converters. Poling pulses were administered across the entire wafer using a TS3000 wafer prober made by MPI, and the electrodes were removed by etching in 1:3 HCl:H$_2$O. Ti/Pt heaters were added after the via etch.


Epitaxy designs for 1030-nm QW, 1150-nm QD and 1180-nm QW epitaxies were bonded and co-processed, the same designs reported in \cite{morin2024coprocessed}, sourced from QD Laser. OFAT active-passive couplers were implemented as in \cite{Morin2024rba}. For the designs reported in this work, the on-chip facets were straight rather than angled, allowing a reflection of approximately 7\%; each gain section was able to lase independently, and the reflections at the facets were intended to provide some longitudinal mode selection.

Triple-gain separated gain PICs were implemented as depicted in Fig. \ref{fig:CONCEPT}c. These PICs had left and right gain sections which were 1200 and 1210 $\mu$m long, respectively; the difference in length was intended to create a Vernier effect between their longitudinal modes and thus reduce the number of lasing modes. Probe-gold rear mirrors were fabricated as described in \cite{Morin2024rba}. At the gold mirrors, the gain sections were nominally 7 $\mu$m wide, but the nominal width was tapered to 2 $\mu$m before the coupers, over a taper length of 400 $\mu$m. The two gain sections on the left side were pumped together using common N- and P-contacts, while the gain section on the right side was controlled separately. Between the left and right gain sections, the PIC consisted of a Y-splitter, a 500-$\mu$m long PPLN section and a directional coupler. The total length of the PIC was <4 mm, and the total footprint was <0.8 mm$^2$.

Unfortunately, during fabrication, insufficient photoresist resulted in waveguides narrower than designed and rough, angled sidewalls. The PPLN quasi-phase-matching was apparently shifted, and waveguide losses were high, especially at the second harmonic wavelengths. Resonator quality factors were spoiled. However, more robust elements such as the OFAT couplers, 100\% directional couplers and Y-splitters were still functional.

\section{Results} \label{sec:RESULT}

In spite of fabrication difficulties, lasers and functional separated gain PICs were demonstrated with all three bonded epitaxies. Steady SHG was visible to the eye, in room lighting, with the best performance measured in multiple separated gain PICs, as depicted in Fig. \ref{fig:CONCEPT}c. 

\begin{figure}[htb]
	\centering
		\includegraphics[width=1.0\textwidth, draft=False]{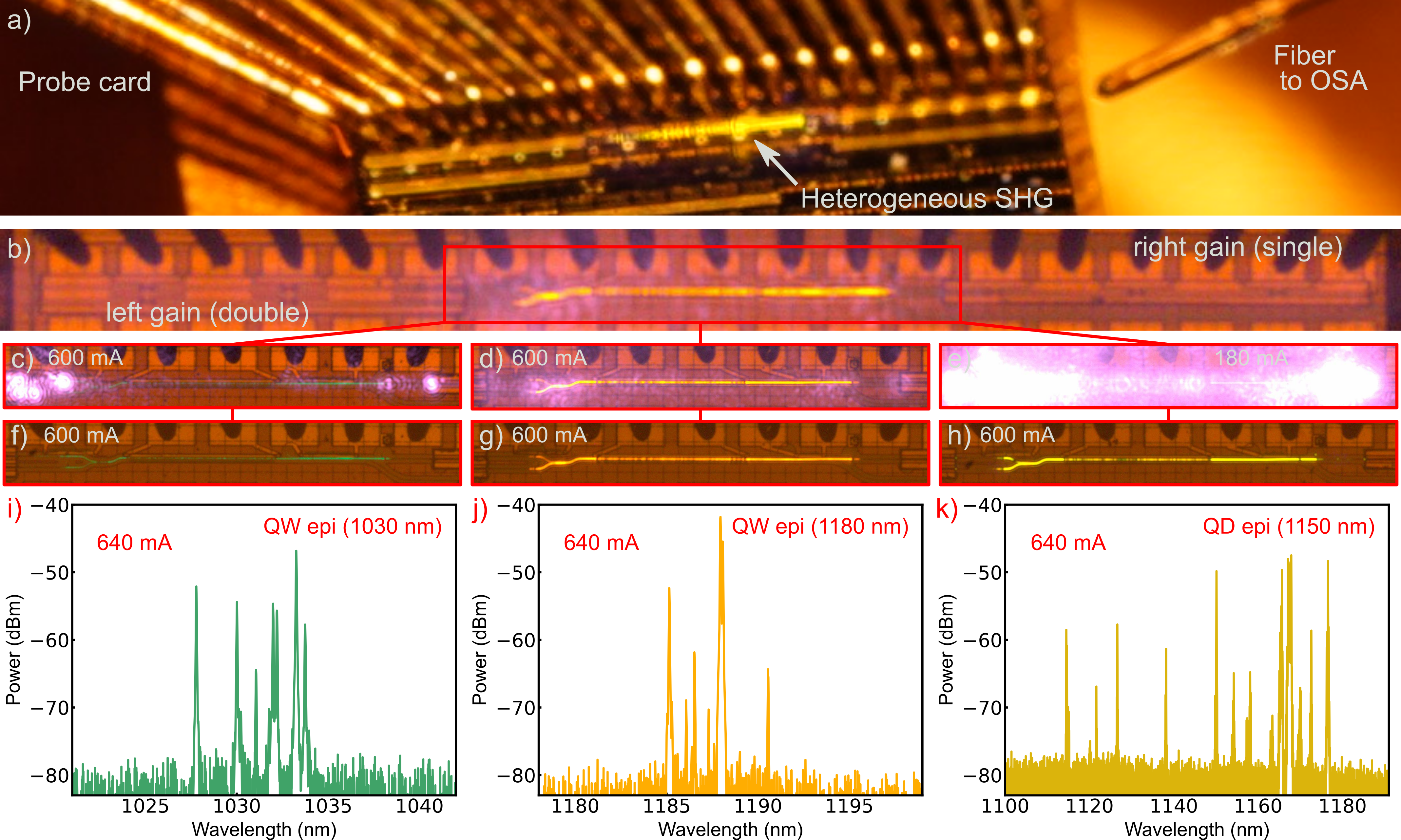}
	\caption[figshorttitle]{Heterogeneously integrated second harmonic generation. (a) DSLR image of second harmonic generation photonic integrated circuit (SHG PIC), Thorlabs FGS900 filter. (b) Microscope image of SHG PIC (no filter). (c-e) Microscope images of SHG PICs with 1030-nm QW, 1180-nm QW and 1150-nm QD epitaxy (no filter). (f-h) The images from (c-d) are repeated with a filter (FGS900). (i-k) Optical spectra of the devices from (c-d).}
	\label{fig:RESULT}
\end{figure}

Figs. \ref{fig:RESULT}a-b shows a full device (DSLR image and microscope image), and Fig. \ref{fig:RESULT}c-h show camera images of the passive sections of different devices in operation, with and without a Thorlabs FGS900 short-pass wavelength filter to remove scattered fundamental light. The contrast of the unfiltered and filtered camera images clearly show that the second harmonic light was produced in the TFLN waveguides between the two gain regions. Notably, the brightest regions are outside of PPLN section, which suggests that a high order mode at the second harmonic frequency offers the best phase-matching to the propagating mode at the fundamental frequency. Because high order modes interact more strongly with the sidewalls, conversion to a high order mode increases the loss contribution from the waveguide etch issues mentioned above.

Propagating in a lossy mode, the SHG light in the output waveguide was insufficient for measurement. However, small amounts of fundamental light escaped the frequency separator. The spectra of this output were measured using a Yokogawa AQ6374 optical spectrum analyzer (OSA), showing that the SHG light must have been produced at wavelengths of roughly 517 nm, 597 nm and 580-590 nm in 1030-nm QW, 1180-nm QW and ``1150-nm'' QD devices, respectively (Fig. \ref{fig:RESULT}i-k). These values agree well with the colors observed by eye and in various cameras. At the highest drive currents, the quantum dot devices lased more strongly in their excited state (in the low 1100's of nm), and the SHG light took on a distinctly greener color.

\begin{figure}[htb]
	\centering
		\includegraphics[width=1.0\textwidth, draft=False]{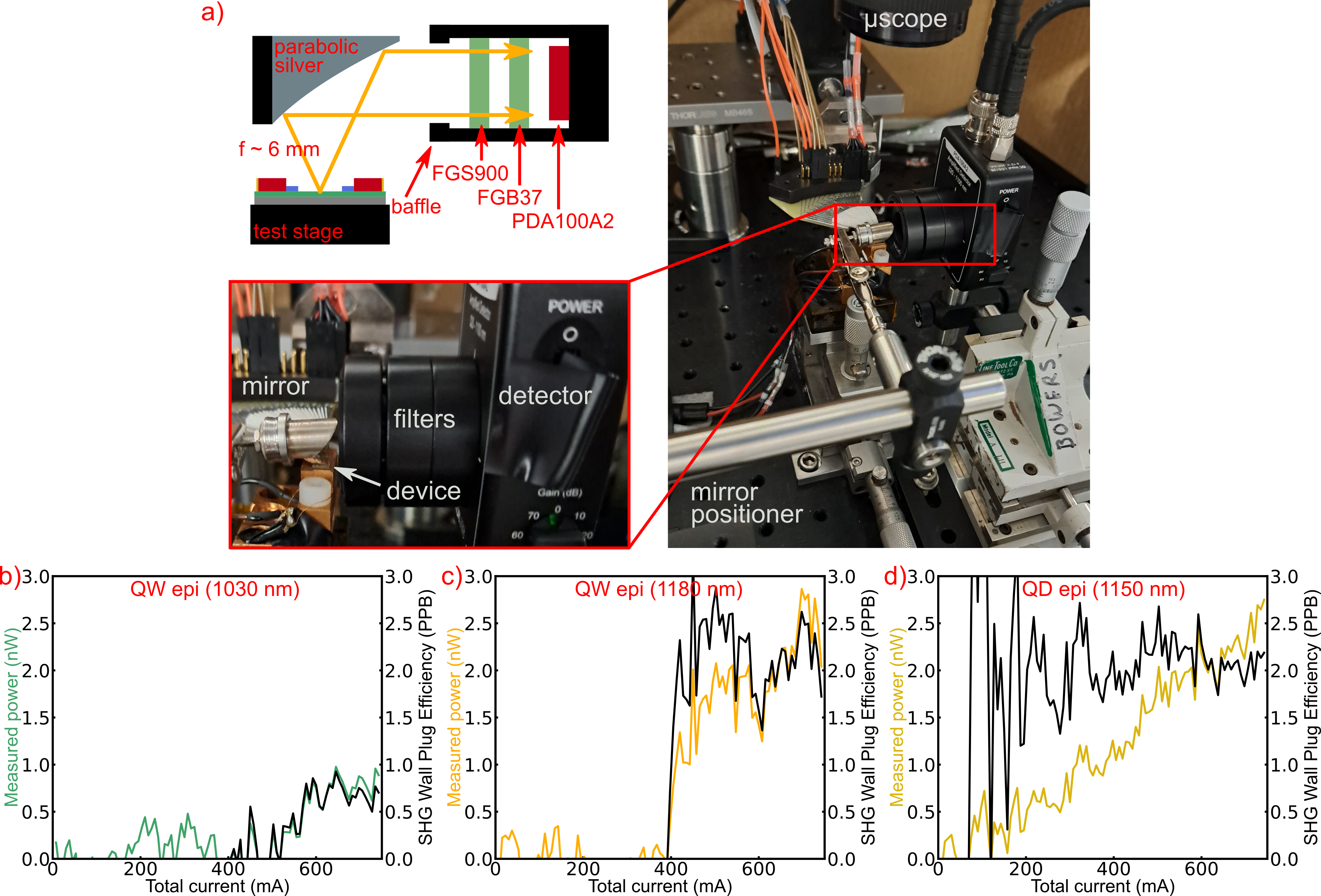}
	\caption[figshorttitle]{Measurement of second harmonic generation photonic integrated circuits (SHG PICs). (a) Measurement setup for measuring scattered SHG. Light is collimated with a parabolic silver mirror and passed through two filters (Thorlabs FGS900 and FGB37) before collection with a Thorlabs PDA100A2 variably amplified photodiode. (b-d) Measured second harmonic power from SHG PICs with 1030-nm QW, 1180-nm QW and 1150-nm QD epitaxy. Second harmonic wall-plug efficiency is plotted in parts per billion.}
	\label{fig:MEASURE}
\end{figure}

SHG power was measured by collimating the upward-scattered light using a silver parabolic mirror. The collimated beam was directed through Thorlabs FGS900 and FGB37 filter glasses and into a PDA100A2 detector at a high amplification setting. This setup is shown in Fig. \ref{fig:MEASURE}a. With this setup, the bias currents of the three gain sections were swept together up to a final combined current of 750 mA, as shown in Figs. \ref{fig:MEASURE}b-d. (Since the two left gain sections shared contacts, they were driven together up to a combined current of 500 mA, while the right gain section was driven up to 250 mA, providing similar current density to all gain material.) The same sweep was executed with and without filter glasses, and the curves in Figs. \ref{fig:MEASURE}b-d are corrected for scattered fundamental light passing through the filters, with attention to the differing responsivities at the different wavelengths. Optical power of >0.5 nW was measured from the 1030-nm QW PIC, while the 1180-nm QW and 1150-nm QD PICs each produced >2 nW, as shown in Figs. \ref{fig:MEASURE}b-d. Considering that much of the scattered light was not directed into the focusing mirror, the generated power is likely higher than these curves indicate. 

Due to the nature of the PIC and device-to-device variation in fabrication, optical conversion efficiency cannot be estimated accurately. However, second harmonic wall plug efficiency (WPE) is plotted in Figs. \ref{fig:MEASURE}b-d. WPE of >2E-9 was demonstrated in the 550-600 nm range. The WPE curves in the better performing devices (Fig. \ref{fig:MEASURE}c and Fig. \ref{fig:MEASURE}d) appear relatively flat (though noisy) rather than increasing. This behavior indicates that the frequency conversion process acted as a linear loss mechanism, implying that frequency conversion was saturated. Since the saturated conversion efficiency declines with phase mismatch \cite{armstrong1962interactions}, the saturation efficiency is expected to improve dramatically with corrected quasi-phase-matching.


\section{Conclusion} \label{sec:CONC}

In this work we have proposed and demonstrated a PIC which promises significant benefits for heterogeneous SHG systems. The PIC was implemented with a flexible GaAs-TFLN platform, and second harmonic light was observed in spite of fabrication difficulties, with power >2 nW and WPE >2E-9.

Far higher efficiency can be achieved in the future with similar designs. Based on laser powers already reported in the heterogeneous GaAs-TFLN platform \cite{morin2024coprocessed} and the conversion efficiency reported in \cite{Wang2018} (a long, single pass periodically poled TFLN device operating at 1500 nm), over 2 mW of SHG power and a second harmonic WPE better than 0.1\% may be achieved by correcting the waveguide process of the current work. In fact, ultra-low loss TFLN waveguides have been demonstrated at visible wavelengths \cite{desiatov2019ultra}, and with resonantly enhanced SHG \cite{Lu2019}, the available power could easily saturate the conversion efficiency. With further improvements to laser efficiency and power output, this approach may yield second harmonic WPE comparable to that of direct emission lasers.

\begin{backmatter}
\bmsection{Funding} This work was funded by the Defense Advanced Research Projects Agency (DARPA) under contract no. HR001-20-2-0044.

\bmsection{Acknowledgments}
The authors would like to thank Joshua Castro and Minh Tran for helpful discussions, as well as Vincent Escueta for assistance with the figures. A portion of this work was performed in the UCSB Nanofabrication Facility, an open access laboratory.

\bmsection{Disclosures}
The authors declare no conflicts of interest.

\bmsection{Data Availability}
Data underlying the results presented in this paper are not publicly available at this time but may be obtained from the authors upon reasonable request.

\end{backmatter}


\bibliography{hshg}

\end{document}